\documentclass{jfm}

\shorttitle{LSC reversals explained by pendulum correspondence}
\shortauthor{N. J. Moore and J. M. Huang}

\title{Large-scale circulation reversals explained by pendulum correspondence}

\author{Nicholas J. Moore\aff{1}
\corresp{\email{nickmoore83@gmail.com}}
  \and Jinzi Mac Huang\aff{2,3} \corresp{\email{machuang@nyu.edu}}}

\affiliation{\aff{1}Department of Mathematics, Colgate University, Hamilton, NY 13346, USA
\aff{2}NYU-ECNU Institute of Physics and Institute of Mathematical Sciences, New York University Shanghai, Shanghai, 200124, China
\aff{3}Applied Math Lab, Courant Institute, New York University, New York, NY 10012, USA}

\usepackage{graphicx, amsmath, amssymb, amsfonts, mathtools, mathrsfs, color}
\usepackage{comment, todonotes}
\usepackage{soul}
\usepackage{dcolumn}

\usepackage{xcolor}
\usepackage[capitalise]{cleveref}
\usepackage[normalem]{ulem}
\usepackage{bm}

\newcommand{\td}[2]{\frac{d #1 }{d #2}}

\newcommand{\pd}[2]{ \frac{ \partial #1}{ \partial #2 } }

\newcommand{\pdi}[2]{ \partial_{#2} #1 }

\newcommand{\bvec}[1]{\ensuremath{\boldsymbol{#1}}}
\newcommand{\grad}{\nabla}

\newcommand{\mean}[1]{ \langle #1 \rangle}

\newcommand{\Pra}{\text{Pr}}
\newcommand{\Ra}{\text{Ra}}

\newcommand{\si}{\text{Supplemental Material}}

\newcommand{\uu}{\bvec{u}}

\newcommand{\Lap}{\grad^2}

\newcommand{\Area}{\text{A}_0}

\newcommand{\len}{h}
\newcommand{\er}{\bvec{e_r}}
\renewcommand{\eth}{\bvec{e_\theta}}

\newcommand{\xc}{X}
\newcommand{\yc}{Y}
\newcommand{\dy}{\Delta y}

\newcommand{\phim}{\phi_{\text{max}}}
\newcommand{\Ebot}{E_{\text{bot}}}
\newcommand{\Lmax}{L_{\text{max}}}
\newcommand{\geff}{g_{\text{eff}}}
\newcommand{\Eeff}{E_{\text{eff}}}

\begin{document}

\maketitle

\begin{abstract}
We introduce a low-order dynamical system to describe thermal convection in an annular domain. The model derives systematically from a Fourier-Laurent truncation of the governing Navier-Stokes Boussinesq equations and accounts for spatial dependence of the flow and temperature fields. Comparison with fully-resolved direct numerical simulations (DNS) shows that the model captures parameter bifurcations and reversals of the large-scale circulation (LSC), including states of (i) steady circulating flow, (ii) chaotic LSC reversals, and (iii) periodic LSC reversals. Casting the system in terms of the fluid's angular momentum and center of mass (CoM) reveals equivalence to a damped pendulum with forcing that raises the CoM above the fulcrum. This formulation offers a transparent mechanism for LSC reversals, namely the inertial overshoot of a forced pendulum, and it yields an explicit formula for the frequency $f^*$ of regular LSC reversals in the high Rayleigh-number limit. This formula is shown to be in excellent agreement with DNS and produces the scaling law $f^* \sim \Ra^{0.5}$.
\end{abstract}

\begin{keywords}
Authors should not enter keywords on the manuscript, as these must be chosen by the author during the online submission process and will then be added during the typesetting process (see http://journals.cambridge.org/data/\linebreak[3]relatedlink/jfm-\linebreak[3]keywords.pdf for the full list)
\end{keywords}

\section{Introduction}
Thermal convection and the associated large-scale circulation (LSC) play an instrumental role in applications diverse as atmospheric and oceanic flows \citep{Salmon1998, Zhong2009}, mantle and liquid-core convection \citep{Whitehead1972, zhang2000periodic, Zhong2005, Whitehead2015, mac2018stochastic}, and solar magneto-hydrodynamics \citep{Wit2020}. In these settings, it is known that the LSC can spontaneously reverse direction \citep{RevModPhys.81.503}, manifesting, for example, as a sudden change in wind direction \citep{Doorn2000} or potentially a reversal of the Earth's magnetic dipole \citep{Glatzmaier1999}.

LSC reversals have been observed in controlled laboratory experiments \citep{Creveling1975, Gorman1984, Gorman1986, Castaing_Gunaratne_Heslot_Kadanoff_Libchaber_Thomae_Wu_Zaleski_Zanetti_1989, Brown2007, Xi2007, Sugiyama2010, Song2011, Wang2018, Chen2019} and numerical simulations \citep{Sugiyama2010, Xu2021}, where a progressive increase of the Rayleigh number Ra triggers a sequence of transitions. Depending on underlying conditions, this sequence can include: (1) the onset of fluid motion giving rise to steady circulation; (2) the destabilization of this circulatory state giving rise to chaotic reversals of the LSC; (3) a return to order at high $\Ra$ in which LSC reversals recur periodically despite small-scale turbulence.

Despite much progress, LSC reversals remain poorly understood. Current theory can be broadly categorized as application of the Lorenz equations or phenomenological models. 
The Lorenz equations, originally derived in the context of planar upper and lower boundaries with unbounded horizontal periodicity \citep{Lorenz1963}, captures many of the transitions listed above. However, when applied to the finite geometries accessible to experiments, the Lorenz system only describes the spatially-averaged flow \citep{Welander1967, Gorman1986, Tritton1988, Widmann1989, Ehrhard1990, Singer1991}, resulting in substantial quantitative differences with experiments \citep{Gorman1986}. More recent phenomenological models account for additional physical effects, such as detached thermal plumes or corner rolls, by supplementing fundamental conservation laws with nonlinear or stochastic terms \citep{Araujo2005, Brown2007, Ni2015}. While these models lend great physical insight, the connection to first principles may not be self-evident due to the {\it ad hoc} nature of the extra terms.
In rectangular and cylindrical domains, it has been suggested that corner vortices and the associated turbulent fluctuations can perturb the LSC structure, causing it to switch between two bistable states \citep{Brown2007, Sugiyama2010}. Switching may occur irregularly, partly due to intermittent heat accumulation and release \citep{Wang2018}.

In this article, we discuss one physical scenario in which a first-principled and precise understanding of LSC reversals can be gained. 
The scenario is thermal convection in a narrow annulus, closely related to a so-called closed-loop thermosyphon \citep{Welander1967, Gorman1986, Tritton1988, Widmann1989, Ehrhard1990, Singer1991, basu2014review}. 
The annular geometry reinforces the dominant circular structure of natural convection, while eliminating corner-induced effects that tend to introduce greater complexity and tend to be geometric specific. The confined nature of the annular flow is more amenable to low-dimensional characterization, while also exhibiting sufficiently complex behavior, such as chaotic and periodic LSC reversals, to be useful for understanding convection. The results of our study complement those conducted in rectangular geometries, where domain corners impact convective behavior \citep{Brown2007, Sugiyama2010, Ni2015, Chen2019}.

Rather than beginning with the Lorenz equations and making adjustments to suit the annulus, we derive a low-order system directly from a Fourier-Laurent truncation of the governing Navier-Stokes Boussinesq (NSB) equations. The resulting system resembles the Lorenz equations, but differs in a few important ways. Notably, the Laurent expansion accounts for spatial dependence of the flow [see also \cite{Yorke1987}], permitting exact enforcement of boundary conditions without the need for empirically estimated friction or heat-transfer coefficients. Comparison to direct numerical simulations (DNS) shows this system captures the entire sequence of transitions, including regimes of chaotic and periodic LSC reversals.

Importantly, we show equivalence between this low-order system and an externally-driven mechanical pendulum. Two lengthscales naturally emerge from this correspondence: the fulcrum $y_1$ of the pendulum and the point $y_0$ towards which the fluid center of mass (CoM) is externally driven, both given by explicit formulas. Knowledge of these quantities yields a simple formula for the frequency of periodic LSC reversals in the high-$\Ra$ regime, shown to be in excellent agreement with DNS. Further, this correspondence reveals a transparent mechanism for reversals, namely the inertial overshoot of the fluid CoM as equivalent to a forced pendulum.
The clean characterization afforded by annular convection may offer a new point of approach for understanding convection in other geometries.

\section{Convective states revealed by DNS}

\begin{figure}
 \includegraphics[width = \textwidth]{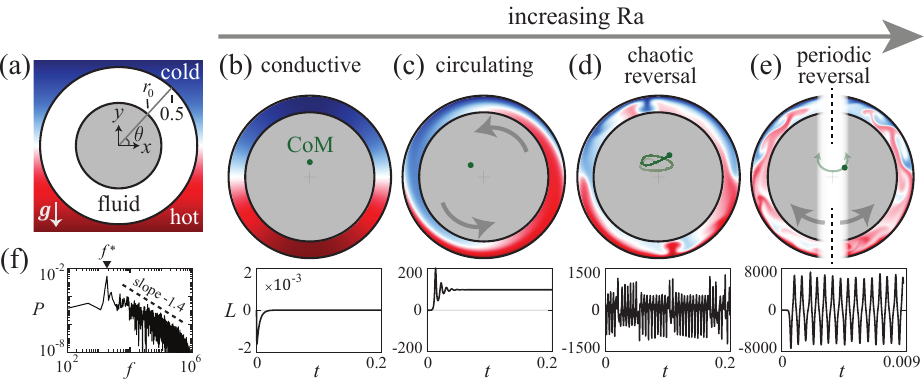}
 \centering
 \caption{Direct numerical simulations of natural convection in an annulus.
 (a) Schematic of an annular fluid domain heated from below.
 (b) At low Ra ($3.9\times10^5$), the conductive state is stable, resulting in a raised CoM (green dot)}. Any initial angular momentum quickly dissipates as shown in the plot of $L(t)$ below. (c) At higher Ra ($3.1\times10^6$) the system transitions to steady circulation with offset CoM and non-zero $L$. (d) At yet higher Ra ($5\times10^7$), the LSC can spontaneously reverse direction. The fluid CoM wanders erratically (green trajectory) and $L(t)$ reverses chaotically.
 (e) At the highest Ra ($1.6\times 10^9$), the LSC reversals recur periodically, even though the small-scale flow is turbulent. (f) The temperature power spectrum of (e) peaks at frequency $f^*$, corresponding to the LSC reversal frequency. At higher frequency, the decay rate is consistent with a $-1.4$ power law. Movies of (b)--(e) are available in \si. 
In all cases, $\Pra = 4$ and $r_0 = 0.4$.

\label{fig1}
\end{figure}

\Cref{fig1}(a) depicts the problem setup in which a 2D annular fluid domain is heated from below. Thermal exchange occurs along the outer boundary with an imposed temperature that decreases linearly with height, while the inner boundary remains adiabatic. Dimensionless temperature $T$, velocity $\uu$, and pressure $p$ fields are governed by the incompressible NSB equations
\begin{align}
\label{NS}
& \pd{\uu}{t} + \uu \cdot \grad \uu = -\grad p +
\Pr \Lap \uu + \Ra\, \Pra \, T \bvec{e_y}, \\
\label{advection}
& \pd{T}{t} + \uu \cdot \grad T = \Lap T, \\
\label{incomp}
&  \grad \cdot \uu = 0 ,
\end{align}
holding in the dimensionless annulus, $r_0<r<1/2$. Both the inner and outer rings are no-slip boundaries. Parameters include the Rayleigh number $\Ra = \beta_T \Delta T \len^3 g/(\nu \kappa)$ and Prandtl number $\Pra = \nu/\kappa$ (\cref{appA}), where $\len{}$ is the dimensional outer boundary diameter and $\Delta T$ is the temperature difference between the bottom and top points of the outer boundary. Other physical parameters include $\beta_T$, $g$, $\nu$, $\kappa$, which are the thermal expansion coefficient, acceleration due to gravity, kinematic viscosity, and thermal diffusivity, respectively. When $\Ra$ is sufficiently high, the destabilizing action of buoyancy can give rise to natural convection.

To quantify different convective states, we will examine the spatially-averaged {fluid angular momentum} $L(t)$ and the fluid CoM coordinates $(\xc(t), \yc(t))$, given by: 
\begin{align}
\label{angularM}
L(t) &= \frac{1}{\Area}\int_{\Omega} r u \, dA , \\
\label{COMx}
\xc(t) &= - \frac{1}{\Area} \int_{\Omega} x \, T \, dA , \\ 
\label{COMy}
\yc(t) &= - \frac{1}{\Area} \int_{\Omega} y \,  T \, dA .
\end{align}
Here, $\Area = \pi(1-4r_0^2)/4$ is the area of the annulus ${\Omega}$ and $dA = r\,dr d\theta$ is the area element. The fluid CoM coordinates above are expressed in terms of the temperature field owing to the fact that fluid density varies as the negative of $T$. We note that $L>0$ corresponds to a counter-clockwise rotating flow.

The range of convective states are revealed by direct numerical simulations (DNS) of the NSB system as shown in \cref{fig1}. Simulations are based on a Chebyshev-Fourier pseudo-spectral discretization (\cref{appB}) of \cref{NS,advection,incomp} in streamfunction-vorticity form with implicit-explicit time stepping \citep{peyret2002spectral, mac2021stable, Huang2022a}.
At low $\Ra$, \cref{fig1}(b) shows the existence of a stable conductive state with no fluid motion and with raised CoM (green dot). In this regime, perturbations to the conductive state  decay rapidly, as seen in the plot below showing $L(t) \to 0$.
Increasing $\Ra$ eventually destabilizes the system, leading to the state shown in \cref{fig1}(c), where the fluid circulates either clockwise (CW) or counterclockwise (CCW) at a constant rate. In this state, the fluid CoM is fixed and offset from center.
By further increasing $\Ra$, this steady circulating state also destabilizes; the direction of circulation now alternates over time and the flow reverses chaotically, as shown in the time series of $L$ in \cref{fig1}(d). The fluid CoM wanders erratically in this regime.
Interestingly, large-scale chaos disappears when $\Ra$ becomes sufficiently high, and \cref{fig1}(e) reveals an oscillating state with periodic LSC reversals. Here, the oscillatory CoM trajectory resembles pendulum motion.
Although the reversals are periodic, the small-scale flow is turbulent and resolved by the DNS. The turbulent fluctuations are characterized by the frequency power spectrum of the temperature field, shown in \cref{fig1}(f) to follow the turbulent Bolgiano-Obukhov power law of natural convection \citep{Wu1990, Lohse2010}.

\section{Low-order model for LSC reversals}

All of these states can be recovered by a low-dimensional system that arises directly from the governing NSB equations. As further detailed in \cref{appC}, the brief derivation is as follows. In polar coordinates, $\uu = u(r,\theta,t) \eth + v(r,\theta,t) \er$ and $T = T(r,\theta,t)$, consider a Fourier expansion in $\theta$ and a Laurent expansion in $r$, and truncate each to a desired order while enforcing all boundary conditions (BCs). The choice of Laurent expansion is guided by the form of the conductive-state solution and recovers this basic state with no approximation made. Inserting the truncated variables into \cref{NS,advection,incomp} and projecting onto the Fourier-Laurent basis yields a finite-dimensional  system. 

Truncating the combined Fourier-Laurent expansion at the lowest order capable of satisyfing all BCs and casting in terms of the physically relevant variables $L(t), \xc(t), \yc(t)$ produces the dynamical system:
\begin{align}
\label{Ldot}
\dot{L} &= - \Ra \Pra \, \xc - \alpha \Pra \, L , \\
\label{xdot}
\dot{\xc} &= -k L (\yc - y_1) - \beta \xc ,  \\
\label{ydot}
\dot{\yc} &= k L \xc - \beta (\yc - y_0) .
\end{align}
This system governs the evolution of the fluid angular momentum $L(t)$ and fluid CoM coordinates $(\xc(t), \yc(t))$ defined in \cref{angularM,COMx,COMy}.
The coefficients $\alpha, \beta, k, y_0, y_1 >0$ are purely geometric in that they depend on $r_0$ only. Formulas for these coefficients are given in \cref{appC}. Though no assumption is made on the width of the annulus, the low-order truncation is most accurate for a relatively narrow annulus. We therefore set $r_0 = 0.4$ in all numerical examples. Higher-order truncations could be used for a wider annulus.

The above system exhibits the same quadratic nonlinearity as the Lorenz equations. The parameter structure, however, arises directly from the annular geometry and differs from that of Lorenz. Differences, therefore, exist in the parameter regimes accessible by each system and especially in the threshold values separating different states.

While the comparison to the Lorenz equations can be illuminating, even more physical insight can be gained by recognizing how \cref{Ldot,xdot,ydot} relate to a mechanical pendulum. In particular, if one artificially sets $\beta = 0$, then \cref{Ldot,xdot,ydot} are {\em identical} to those of a pendulum with fulcrum $y_1$, length $l = \sqrt{\xc^2 + (\yc-y_1)^2}$, effective gravitational constant $\geff = k l^2 \, \Ra \, \Pra$, and damping rate $\alpha \Pra$. The system is simply written in terms of the pendulum CoM and angular momentum rather than the more familiar angular displacement. 
In this equivalence, the pendulum CoM corresponds exactly to the fluid CoM $(\xc,\yc)$ as defined in \cref{COMx,COMy}, and the pendulum angular momentum corresponds exactly to the fluid angular momentum $L$ as defined in \cref{angularM}. Note that the effective gravitational constant $\geff$ of the pendulum system is unrelated to the actual gravitational constant $g$ of the thermal convection system. Also note that if $\beta \ne 0$, the two additional driving terms present in \cref{xdot,ydot} can cause the pendulum length $l = \sqrt{\xc^2 + (\yc-y_1)^2}$ to vary dynamically, opening the possibility of chaotic dynamics.

The driving terms involving $\beta$ arise from the interaction of boundary heating and buoyancy. As seen in \cref{Ldot,xdot,ydot}, these terms drive the CoM towards the point $(0,y_0)$, which \cref{appA} shows is the CoM of the conductive state.
Thus, $(L,\xc,\yc)=(0,0,y_0)$ corresponds to the conductive-state solution that is given explicitly by \cref{conductive} and depicted in \cref{fig1}(b). Stability analysis discussed in \cref{Sec:bifurcations} shows that this state is stable up to a critical Rayleigh number.

The most important parameters in the pendulum correspondence of \cref{Ldot,xdot,ydot} are $y_0$ and $y_1$, corresponding to the height of the conductive-state CoM and the pendulum fulcrum respectively. \cref{appA,appC} give explicit formulas for these two length scales in terms of the geometric parameter $r_0$.
When the fluid CoM lies at the pendulum fulcrum, $(\xc,\yc) = (0,y_1)$, the restoring torque in \cref{Ldot,xdot,ydot} vanishes but the driving terms that push $(\xc,\yc)$ towards $(0,y_0)$ do not vanish. The state $(L,\xc,\yc) = (0,0,y_1)$ is therefore not an equilibrium of the system due to the continual injection of thermal energy from the boundary. \Cref{appC} further shows that $y_0 > y_1 > 0$ for any choice of $r_0$, implying that the thermal injection  always acts to raise the CoM above the fulcrum and, hence, tends to destabilize the system.

\section{Bifurcations and comparison with DNS}
\label{Sec:bifurcations}

\begin{figure}
\includegraphics[width=0.85\textwidth]{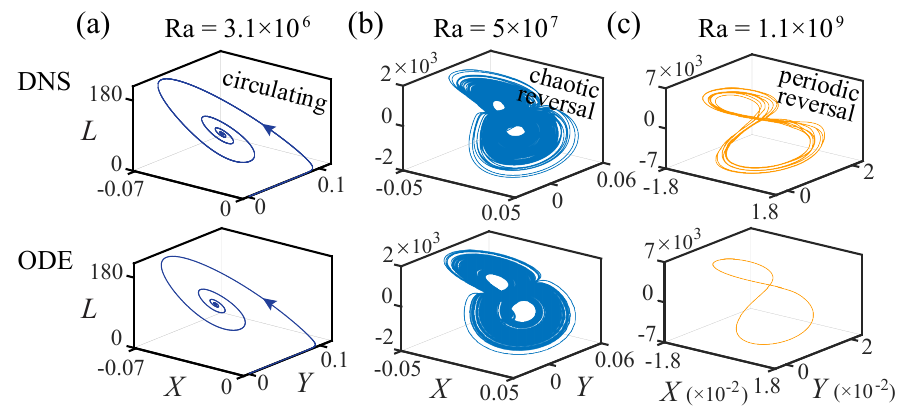}
 \centering
\caption{Trajectories of ODE system \eqref{Ldot}--\eqref{ydot} in comparison to fully-resolved DNS. The trajectories of $(L,\xc,\yc)$ are remarkably similar across the range of Rayleigh numbers, showing (a) convergence to a stable circulating state for $\Ra = 3.1\times10^6$, (b) strange-attractor dynamics for $\Ra = 5\times10^7$, and (c) periodic dynamics for $\Ra = 1.1\times10^9$.
In all cases, $\Pra = 4$ and $r_0 = 0.4$.}
\label{fig2}
\end{figure}

How well does this simple ODE system describe the dynamics of convection? \Cref{fig2} shows trajectories of $(L,\xc,\yc)$ computed by fully-resolved DNS (top) versus those computed by the ODE model (bottom) for the same Rayleigh numbers as \cref{fig1}(c)--(e). \Cref{fig2}(a)--(c) shows that the trajectories from DNS and the ODE model are remarkably similar across the range of $\Ra$, exhibiting (a) convergence to a stable circulating state, (b) chaotic dynamics near a strange attractor, and (c) periodic orbits at the highest $\Ra$. The trajectories in \cref{fig2}(b)-(c) indicate reversals of the LSC, as can be seen by the sign change of $L$. The LSC reversals are chaotic in \cref{fig2}(b) and periodic in \cref{fig2}(c).

\begin{figure}
\includegraphics[width=0.6\textwidth]{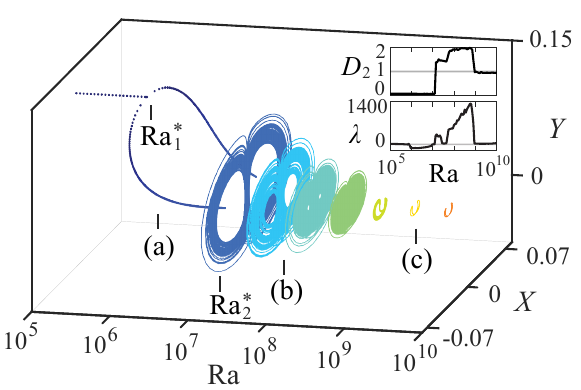}
 \centering
\caption{Bifurcation diagram shows a pitchfork bifurcation at $\Ra_1^*$ and a Hopf bifurcation at $\Ra_2^*$. Trajectories corresponding to \cref{fig2}(a)-(c) are marked on the diagram; For all trajectories, $\Pra = 4$ and $r_0 = 0.4$.
\textit{Inset:} The fractal dimension $D_2$ and Lyapunov exponent $\lambda$ distinguish chaotic states from orderly ones. }
\label{fig3}
\end{figure}

The bifurcation diagram in \cref{fig3} shows that a pitchfork bifurcation occurs at a critical value $\Ra_1^*$. At this value, the conductive state loses stability, and, simultaneously, the bistable circulating states appear (CW and CCW circulation). 
At a second critical value, $\Ra_2^*$, these circulating states lose stability through a Hopf bifurcation. Immediately past $\Ra_2^*$, the dynamics are fractal-like and chaotic, characteristic of a strange attractor. These observations are further supported by measurements of the fractal dimension $D_2$ \citep{ott2002chaos} and Lyapunov exponent $\lambda$ shown in the inset. At much higher $\Ra$, order reemerges and trajectories resemble the arc-like path of a pendulum.

The ODE model yields exact formulas for both critical values (\cref{appD}):
\begin{equation}
\label{ra12}
\Ra_1^*  =  \frac{\alpha\beta}{k\dy}, \quad
\Ra_2^* = \frac{\alpha^2 \, \Pra}{k \dy} \left( 
\frac{\alpha \Pra + 4\beta}{\alpha \Pra - 2\beta} 
\right),
\end{equation}
where $\dy = y_0-y_1>0$ is the distance between the conductive-state CoM and the pendulum fulcrum. Briefly, the value $\Ra_1^*$ is found through linear stability analysis of the conductive state $(L,\xc,\yc) = (0,0,y_0)$. As $\Ra$ crosses $\Ra_1^*$, the conductive state loses stability and the circulating states appear.
Immediately past $\Ra_1^*$, the Jacobian of each circulating state possesses three real, negative eigenvalues. As $\Ra$ increases further, two eigenvalues become complex, $z_{2,3} = \sigma \pm i \omega$, with $\sigma<0$ initially. As $\Ra$ crosses $\Ra_2^*$, $\sigma$ becomes positive and thus the circulating states lose stability, giving way to the strange attractor seen in \cref{fig2}(b). 
This analysis is similar to that conducted for the Lorenz equations \citep{Welander1967, Creveling1975, Gorman1986, Ehrhard1990}, the main difference being that modelling choices made early on (e.g.~accounting for the flow's spatial dependence) enable greater accuracy in predicting the bifurcation parameters than obtained previously \citep{Gorman1986}.

\begin{figure}
 \includegraphics[width=0.6\textwidth]{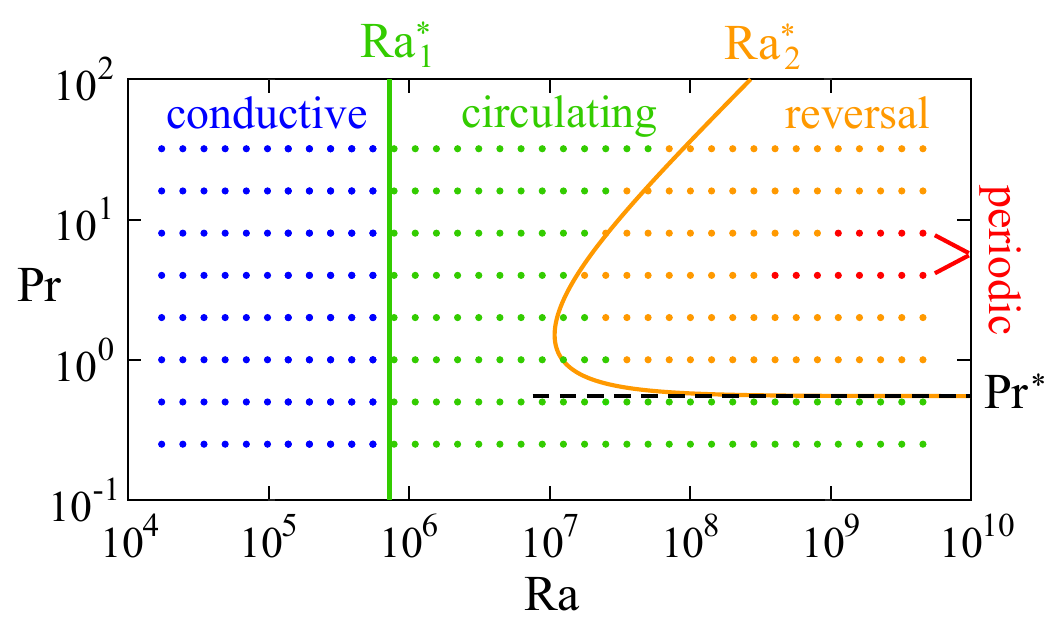}
 \centering
  \caption{Phase diagram of different convective states. Colored dots are from DNS, where blue indicates a stable conductive state, green indicates bistable circulating states, orange indicates chaotic LSC reversals, and red indicates periodic LSC reversals.
  Formulas for $\Ra_1^*$ and $\Ra_2^*$ from the ODE model predict the boundaries between the regions well.
  }
\label{fig4}
\end{figure}

The phase diagram in \cref{fig4} gives a bird's-eye view of the different convective states. In the figure, colored dots correspond to fully-resolved DNS, showing regions of a stable conductive state (blue), bistable circulating states (green), and LSC reversals, both chaotic (orange) and periodic (red). The boundaries between these regions are well predicted by the formulas for $\Ra_1^*$ and $\Ra_2^*$ given in \cref{ra12}. In particular, $\Ra_1^*$ is independent of the Prandtl number, giving the vertical green line. The predicted value of $\Ra_1^*$ agrees with DNS in all cases to within the grid resolution of \cref{fig4}. Meanwhile, the orange curve shows the $\Pra$ dependence of $\Ra_2^*$. For the important case of water, $4 < \Pra < 8$, this threshold is also predicted to within the grid resolution. While some discrepancy is visible for other values of $\Pra$, the curve captures the qualitative shape of the boundary, in particular the horizontal asymptote $\Pra^* = 2 \beta/\alpha$ obtained by setting the denominator of $\Ra_2^*$ equal to zero.
We note that, even though it is a very different geometry, thermal convection in a rectangular domain yields a phase diagram with the same states and with similar orders of magnitude for the thresholds \citep{Araujo2005}.

\section{Periodic LSC reversals at high $\Ra$}

As $\Ra$ increases well beyond $\Ra_2^*$, large-scale chaos subsides and gives way to the nearly periodic trajectories seen in \cref{fig2}(c). While the bifurcations discussed in the previous section have been qualitatively described by related models, the periodic regime has received less attention and, so far, has resisted clean characterization. It is precisely this regime where the novel mechanical-pendulum correspondence becomes most valuable.

\begin{figure}
\includegraphics[width=0.6\textwidth]{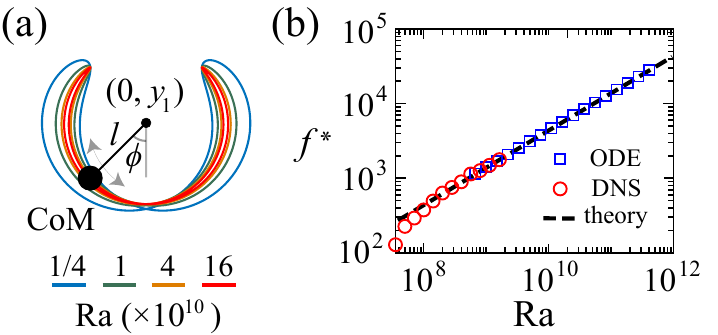}
\centering
\caption{
At very high $\Ra$, order reemerges and large-scale dynamics become periodic. (a) The fluid CoM follows the swinging motion of a pendulum about fulcrum $(0,y_1)$. (b) The  frequency of LSC reversals is well predicted by \cref{f*} for high $\Ra$. In all cases, $\Pra = 4$ and $r_0 = 0.4$.
}
\label{fig5}
\end{figure}

As seen in \cref{fig3} inset, the return to order at high $\Ra$ is indicated by the fractal dimension dropping to one and the Lyapunov exponent dropping to zero at the same Rayleigh number, roughly $\Ra = 10^9$. At this value, a stable limit cycle emerges in the ODE system, giving CoM orbits that resemble pendulum motion. \Cref{fig5}(a) shows  four such orbits of the fluid CoM $(\xc(t),\yc(t))$ for Rayleigh numbers in the range  $1/4 \times 10^{10} < \Ra < 16 \times 10^{10}$. At the lowest $\Ra$, the corresponding pendulum length $l(t) = \sqrt{\xc^2 + (\yc-y_1)^2}$ varies somewhat over the period. At higher $\Ra$, though, the orbit tightens and $l$ remains nearly constant throughout.
Recall that, even though the large-scale dynamics of the fluid angular momentum $L(t)$ and CoM $(\xc(t),\yc(t))$ are regular in this regime, the DNS shows that turbulent fluctuations inhabit the small scales [see \cref{fig1}(e)]. 

Each swing of the pendulum seen in \cref{fig5}(a) corresponds to a sign-change of the fluid angular momentum $L$ and, therefore, a reversal of the large-scale circulation. This simple observation offers a way to predict the dominant frequency $f^*$ of LSC reversals. In the pendulum correspondence of \cref{Ldot,xdot,ydot}, the gravitational constant is $\geff = k l^2 \, \Ra \, \Pra$, giving a small-amplitude frequency of $\sqrt{k l \, \Ra \, \Pra}/(2\pi)$. The amplitudes seen in \cref{fig5}(a), however, are not small, implying that the frequency depends on both the pendulum length $l$ and the maximum swing angle $\phim$. 

As detailed in the \cref{appE}, both of these quantities can be estimated through an energy balance with $\Eeff = \frac{1}{2} k L^2 + \Ra\, \Pra\, (\yc-y_1)$ representing the sum of kinetic and potential energy of the mechanical pendulum system. This quantity is an effective energy of the pendulum system and does not directly represent the actual energy of the thermal convection system.
Although $\Eeff$ is not necessarily conserved over the dynamics, it does satisfy a precise energy law with energy injection (from boundary heating) and dissipation (from fluid viscosity). In the case of a limit cycle, the total energy injected must balance that dissipated, and the period-averaged $\Eeff$ is conserved. 
With a few additional approximations made, this principle allows one to solve for pendulum length and maximum swing angle that arise in the case of a limit cycle. The resulting values of $l$ and $\phim$ depend on geometry and $\Pra$, but not on $\Ra$ (see \cref{appE}). With these values known, the (dimensionless) frequency of high-$\Ra$ LSC reversals is given by
\begin{equation}
\label{f*}
f^* = \frac{\sqrt{kl \, \Ra \, \Pra\,}}{ 4 \, K(\sin^2 ({\phim}/{2})) },
\end{equation}
where $K$ is the complete elliptic integral of the first kind.

\Cref{fig5}(b) shows a comparison between this simple formula and the reversal frequency measured in the fully-resolved DNS. In the DNS, the reversal frequency is measured as the peak location in the temperature power spectrum [see \cref{fig1}(f)]. \Cref{fig5}(b) shows that \Cref{f*} predicts the reversal frequency measured in DNS remarkably well over the largest decade of $\Ra$ run (roughly $\Ra = 2\times 10^8$ to $2\times10^9$). At higher $\Ra$, DNS becomes computationally prohibitive but numerical solution of the ODE model is feasible, and the corresponding measurements of $f^*$ also agree with \cref{f*}. The close agreement between DNS, the ODE model, and \cref{f*} suggests the primary mechanism for LSC reversals has been properly accounted for by the mechanical pendulum correspondence.

In dimensional terms, \cref{f*} gives a reversal frequency of $F^* = c N^*$, where $N^* = \sqrt{\beta_T \Delta Tg/h}$ is the Brunt–V\"ais\"al\"a frequency, i.e.~the inverse of the free-fall timescale, and $c$ is a constant that  depends on geometry and $\Pra$, but not $\Ra$ (\cref{appE}). We note that $c=0.04$ for the case of \cref{fig5}, indicating roughly 25 free-fall timescales per reversal event. Other geometries may yield different values of this ratio.

\section{Discussion}

The low-order system given by \cref{Ldot,xdot,ydot} arises from systematic analysis of the governing NSB equations and has been shown to accurately describe a range of convective states in the annular domain. In contrast with related Lorenz-type models \citep{Welander1967, Gorman1986, Tritton1988, Widmann1989, Ehrhard1990, Singer1991,Araujo2005}, the Laurent expansion underlying \cref{Ldot,xdot,ydot} accounts for spatial dependence of the flow and temperature fields, precluding the need for empirically estimated friction or heat-transfer coefficients. This modeling choice enables greater accuracy in predicting parameter bifurcations, as demonstrated by direct comparison with fully-resolved DNS. 
Many of these related models have been used as the foundation for control \citep{Singer1991, Singer1992}, data assimilation \citep{Harris2012, chen2018conditional}, and machine learning \citep{chen2020learning}. The accuracy and conceptually transparency afforded by \cref{Ldot,xdot,ydot} could further such endeavors.

Importantly, this low-order system reveals a previously unrecognized pendulum structure underlying natural convection. In particular, \cref{Ldot,xdot,ydot} correspond to a damped pendulum with CoM driven upwards towards the conductive-state CoM. In addition to its physical elegance, this equivalence enables accurate predictions for the frequency of regular LSC reversals observed at high $\Ra$. Furthermore, it provides a transparent mechanism for the reversals. Just like a mechanical pendulum, inertia causes the fluid CoM to overshoot equilibrium. The CoM eventual reaches a zenith, at which point the restoring torque reverses the system's angular momentum, thereby creating a LSC reversal. The driving terms in \cref{Ldot,xdot,ydot} are necessary to counteract the damping from viscous dissipation; it is the interplay between these two terms that selects the effective pendulum length, the swing angle, and thus the frequency of reversals.

This low-order model is closely related to the Lorenz equations; indeed, a change of variables can map \cref{Ldot,xdot,ydot} to a system with the same variable structure but a different parameter structure as Lorenz. A corollary of this fact is that the Lorenz system too could be usefully regarded as an externally-driven mechanical pendulum. This observation may lend new insights into the study of the Lorenz equations. We note that, although periodic solutions of the Lorenz equations have been explored mathematically \citep{robbins1979periodic, Sparrow2012}, the physical connection to LSC reversals was not made.

A benefit of the annular domain is that it accentuates the dominant circulating pattern of natural convection, while suppressing other, geometric-specific effects. Related to this fact, we have found that the annulus yields a simple law for the $\Ra$-dependence of the periodic LSC reversal frequency $f^*$, described by an explicit formula, \cref{f*}, and corroborated by comparison with DNS. Previous theoretical analysis in a rectangular geometry suggests the power law $f^* \sim \Ra^{0.44}$ \citep{Araujo2005}, and laboratory experiments with cryogenic helium gas in a cylindrical container suggest $f^* \sim \Ra^{0.71}$ \citep{Araujo2005}.
In the present case of an annulus, the modeling prediction and DNS are in agreement, both unambiguously showing a scaling of $f^* \sim \Ra^{0.5}$. 
Therefore, annular convection may be considered an ideal `ground-state' that yields a precisely determined scaling law for the frequency of reversals. 
Perhaps future studies could build upon this model to determine how the scaling law is modified by various geometric effects.

Curiously, the scaling law $f^* \sim \Ra^{0.5}$ is seen in experimental measurements of thermal convection in a disk \citep{Song2011}, but for the frequency of oscillations in the strength of the LSC. The period of this oscillation is much shorter than the average reversal time seen in the experiments \citep{Wang2018}, suggesting differences in the LSC reversals that occur in disk convection.
In the annular domain, the inner boundary at $r = r_0$ serves as a confinement that regulates the flow. The recent study of \cite{Li2024} demonstrates this concept experimentally, as the inclusion of a central obstruction in Rayleigh-B\'enard convection substantially modifies the flow structures and enhances heat transfer.

Here, we have focused on the lowest-order system capable of satisfying the BCs on the annulus, but the truncation procedure can in principle be carried out to any order. At extremely high Ra, turbulent effects \citep{Lohse2010} are associated with fine-scale structures, which could potentially be captured by retaining higher-order terms in the Fourier-Laurent expansion, either directly or through stochastic parameterization.
Moreover, extension of the model into three dimensions could account for azimuthal rotations of the LSC plane, which experiments have shown take a stochastic character \citep{Brown2005}. Finally, we hope to couple the model to slowly-moving boundaries to examine phase-change processes, such as melting or dissolution, that couple to the action of natural convection \citep{Huang2015,Moore2017,Huang2020a,Huang2022,Weady2022}.

\section*{Funding}
The authors thank Jun Zhang for useful discussions. J.~M.~Huang acknowledges support from the National Natural Science Foundation of China (12272237, 92252204). N.~J.~Moore acknowledges support from the National Science Foundation (DMS-2012560).

\section*{Declaration of interests}
The authors report no conflict of interest.

\section*{Author contributions}
N.~J.~Moore and J.~M.~Huang contributed equally to this work.

\appendix
\section{Formulation and background}\label{appA}
To obtain the dimensionless \cref{NS,advection,incomp}, space has been rescaled on $\len$ (the diameter of the outer annulus boundary), time on $\len^2 / \kappa$ (the thermal diffusive timescale), velocity on $\kappa / \len$, and density variations on $\Delta \rho = \rho_0 \beta_T \Delta T$. 
The prescribed temperature on the outer boundary, $r=1/2$, decreases linearly with height, while the inner boundary, $r=r_0$, remains adiabatic. The velocity field, expressed as $\uu = u \eth + v \er$ in polar coordinates, satisfies no-slip conditions on both boundaries. The boundary conditions (BCs) are thus:
\begin{align}
\label{noslip}
&u = v = 0 \hspace{25pt} \text{at } r=r_0 \text{ and } r=1/2, \\
\label{Tinner}
&\pd{T}{r} = 0 \hspace{35pt} \text{at } r=r_0, \\
\label{Touter}
&T = \frac{1-\sin \theta}{2} \hspace{10pt} \text{at } r=1/2 .
\end{align}

\Cref{NS,advection,incomp} support a conductive-state solution with no fluid motion $(u,v) = (0,0)$. The corresponding temperature field that satisfies BCs \eqref{Tinner}--\eqref{Touter} is given by
\begin{align}
\label{conductive}
T = \frac{1}{2} - &\frac{r_0}{1+4r_0^2} \left(\frac{r}{r_0}+\frac{r_0}{r}\right)\sin{\theta}.
\end{align}

The fluid angular moment $L$ and fluid CoM coordinates (see \cref{angularM,COMx,COMy}) associated with the conductive-state solution are
\begin{align}
L = 0, \quad \xc = 0, \quad \yc = y_0 = \frac{ 1+12 r_0^2 }{16 (1+4r_0^2)}.
\end{align}
That is, $y_0$ represents the CoM of the conductive state. As seen above, $y_0 > 0$ for any value of $r_0$, indicating that the conductive-state CoM always lies above the annulus center.

\section{Numerical methods}\label{appB}

\Cref{NS,advection,incomp} can be written in the streamfunction-vorticity form:
\begin{align}
\label{omegaeq}
\pd{\omega}{t} + &\uu \cdot \grad \omega = \Pra \Lap \omega   + \Pra\, \Ra\, \left(\frac{\partial T}{\partial r}\cos\theta-\frac{1}{r}\frac{\partial T}{\partial \theta}\sin\theta\right),\\
\label{Teq}
&\pd{T}{t} + \uu \cdot \grad T = \Lap T,\\
\label{psieq}
 -&\Lap \psi = \omega,\ \  \uu = \nabla_\perp \psi.
\end{align}
Where the vorticity is $\omega = r^{-1} \left[\partial_r(r u)-\partial_\theta v\right]$, and the stream function $\psi$ recovers the flow velocity through $\uu  = \nabla_\perp \psi = r^{-1}\psi_\theta\er - \psi_r\eth$. 

Discretizing time with the second-order Adam-Bashforth Backward Differentiation method (ABBD2), \cref{omegaeq,Teq,psieq} become
\begin{align}
\label{omegadisc}
\Lap \omega^{(n)} -\sigma_1\omega^{(n)}  &= f^{(n)},\\
\label{Tdisc}
\Lap T^{(n)} -\sigma_2 T^{(n)}  &= g^{(n)},\\
\label{psidisc}
 -\Lap \psi^{(n)} &= \omega^{(n)},
\end{align}
at time step $t = n\Delta t$. Here
\begin{align}
\Lap &= \frac{\partial^2}{\partial r^2} + \frac{1}{r}\pd{}{r} + \frac{1}{r^2} \frac{\partial^2}{\partial\theta^2}, \\
\sigma_1 &= \frac{3}{2\,\Pra\,\Delta t},\quad \sigma_2 = \frac{3}{2\Delta t}, \\[8pt]
f^{(n)} = \Pra^{-1}&\left[ 2  (\uu \cdot \grad \omega)^{(n-1)} - (\uu \cdot \grad \omega)^{(n-2)}\right] \\
&- (2\, \Pra\, \Delta t)^{-1} \left(4\omega^{(n-1)}-\omega^{(n-2)}\right) \notag \\
&-\Ra\, \left(T_r\cos\theta-r^{-1}T_\theta\sin\theta\right)^{(n)},\notag \\[8pt]
g^{(n)} = &\left[ 2  (\uu \cdot \grad T)^{(n-1)} - (\uu \cdot \grad T)^{(n-2)}\right]  \\
&- (2\Delta t)^{-1} \left(4T^{(n-1)}-T^{(n-2)}\right) .\notag
\end{align}

Explicit and nonlinear terms in $f^{(n)}$ and $g^{(n)}$ are computed pseudo-spectrally with an efficient anti-aliasing filter \citep{Hou2007}. \Cref{omegadisc,Tdisc,psidisc} are then solved by Fourier-Chebyshev method detailed in \cite{peyret2002spectral,Huang2022a,PRF2023}. There are typically 1024 Fourier modes and 128 Chebyshev nodes in each DNS of this article. The time step is $\Delta t = 5\times 10^{-4} \,\Ra^{-1/2}$, considering the flow velocity $|\uu|\sim \sqrt{\Ra}$. These parameters are tested to yield resolved and accurate solutions. 
For the ODE model, we use the \textit{ode45} package of MATLAB. 

\begin{figure}
\includegraphics[width=0.6\textwidth]{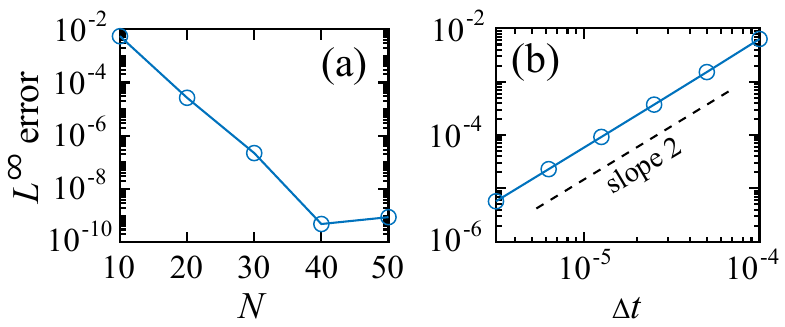}
 \centering
\caption{Convergence of the numerical solver. (a) Spatial convergence test shows the error decays spectrally. (b) Temporal convergence test demonstrates a second order convergence in time stepping. }
\label{fig6}
\end{figure}

\Cref{fig6} shows the convergence test of the DNS scheme. In the spatial convergence test \cref{fig6}(a), the time step $\Delta t = 10^{-4}$ is fixed and $N_r = N_\theta = N$. A high-resolution solution with $N = 1024$ is computed first and the convergence towards this solution is tested by letting $N$ progressively increase. \Cref{fig6}(a) shows the results of this test and demonstrates {\em spectral convergence}. That is, the error decreases exponentially with $N$ until a limiting error of roughly $10^{-9}$ is reached. In the temporal convergence test shown in \cref{fig6}(b), $N_r = N_\theta = 100$ is fixed and $\Delta t$ is decreased by half during each test and then the error between each refinement is compared. Shown in \cref{fig6}(b), the refinement error decays as $\mathcal{O}(\Delta t^2)$, demonstrating a second order temporal convergence.

\section{Derivation of the ODE model}\label{appC}

In polar coordinates, the $\theta$ components of \cref{NS,advection,incomp} are given by
\begin{align}
\label{NSeth}
u_t + v u_r + &\frac{1}{r} u u_\theta +  \frac{1}{r} u v = -\frac{1}{r} p_\theta + \Ra \Pra \, T \cos \theta \\
&+\Pra \left( u_{rr} + \frac{1}{r} u_r + \frac{1}{r^2} u_{\theta \theta} - \frac{1}{r^2} u + \frac{2}{r^2} v_\theta \right) \, , \notag \\[8pt]
\label{HeatTrans}
T_t + \frac{u}{r} T_\theta &+ v T_r = \frac{1}{r} \pd{}{r} \left( r T_r \right) + \frac{1}{r^2} T_{\theta \theta} \, ,\\[8pt]
\label{incomp2}
v_r + \frac{1}{r} v +& \frac{1}{r} u_\theta = 0 \, .
\end{align}
Multiplying \cref{NSeth} by $r^2$, integrating over the fluid domain, applying incompressibility \eqref{incomp2} and the no-slip condition \eqref{noslip}, and using the CoM definition \eqref{COMx}--\eqref{COMy} gives 
\begin{equation}
\label{Ldot1}
\dot{L} = -\Ra\, \Pra \, \xc + 
\frac{\Pra}{\Area} \int_0^{2\pi} \left(r^2 u_r \right)\Big|_{r_0}^{r_1} \, d\theta.
\end{equation}
This evolution equation for the fluid angular momentum is exact within the NSB framework.

The temperature and velocity fields $T(r,\theta,t)$, $u(r,\theta,t)$, $v(r,\theta,t)$ are each periodic in $\theta$ and so can be written as Fourier series,
\begin{align}
\label{TFourier}
T(r,\theta,t) &= a_0(r,t) + \sum_{n=1}^{\infty}  a_n(r,t) \cos n\theta + b_n(r,t) \sin n\theta ,\\
\label{uvFourier}
u(r,\theta,t) &= \sum_{n=-\infty}^{\infty} \hat{u}_n(r,t) e^{i n \theta},\\
v(r,\theta,t) &= \sum_{n=-\infty}^{\infty} \hat{v}_n(r,t) e^{i n \theta} .\notag
\end{align}
The temperature BCs \eqref{Tinner}--\eqref{Touter} imply
\begin{align}
\label{abinner}
& \pdi{a_n}{r} = \pdi{b_n}{r} = 0 \quad \text{ at } r=r_0, \\
\label{abouter}
& a_0 = 1/2, b_1 = -1/2, \text{  all others vanish at } r=1/2.
\end{align}
The no-slip BC \eqref{noslip} and incompressibility \eqref{incomp2} yield conditions
\begin{align}
\label{uvh_bc}
&\hat{u}_n(r,t)=\hat{v}_n(r,t) = 0
\quad \text{at } r=r_0 \text{ and } r=1/2, \\
\label{uvh_incomp}
&i n \hat{u}_n + \hat{v}_n + r \pdi{\hat{v}_n}{r} = 0 \quad \text{for } r\in(r_0,1/2) .
\end{align}

Given the constraints \cref{uvh_bc,uvh_incomp}, the lowest-order truncation of \cref{TFourier,uvFourier} possible is
\begin{align}
    \label{truncuv}
    &u(r,t) = \hat{u}_0(r,t),\quad v(r,t) = 0,\\
    \label{truncTFourier}
    &T(r,\theta,t) = a_0(r,t) +  a_1(r,t) \cos \theta + b_1(r,t) \sin \theta ,
\end{align}
where $a_0, a_1, b_1, \hat{u}_0$ satisfy the boundary conditions \cref{abinner,abouter,uvh_bc}.

Inserting \cref{truncuv,truncTFourier} into \cref{HeatTrans}, multiplying by $r^2$ and projecting onto respective Fourier mode gives
\begin{align}
    \label{a0}
    &\dot{a}_0 = r^{-1} \pdi{}{r} \left( r \pdi{}{r} a_0 \right) ,\\
    \label{andot}
    & r^2 \dot{a}_1 = - r \, \hat{u}_0\, b_1 - a_1 + r \pdi{}{r} \left( r \pdi{}{r} a_1 \right) , \\
    \label{bndot}
    & r^2 \dot{b}_1 = + r \, \hat{u}_0\, a_1 - b_1 + r \pdi{}{r} \left( r \pdi{}{r} b_1 \right) .
\end{align}
Boundary conditions \eqref{abinner}--\eqref{abouter} then imply $\lim_{t\to \infty} a_0(r,t) = 1/2$ irrespective of the initial condition.
We therefore set $a_0 = 1/2$, since variations from this value simply reflect transient dynamics that are decoupled from the rest of the system.

From \cref{angularM,COMx,COMy}, $L,\, \xc,\, \yc$ can be evaluated as
\begin{align}
\label{L_int}
L(t) &= \frac{2\pi}{\Area}\int_{r_0}^{1/2} r^2 \hat{u}_0(r,t) \, dr ,\\
\label{x_int}
\xc(t) &= -\frac{\pi}{\Area} \int_{r_0}^{1/2} r^2 a_1(r,t) \, dr , \\
\label{y_int}
\yc(t) &= -\frac{\pi}{\Area} \int_{r_0}^{1/2} r^2 b_1(r,t) \, dr .
\end{align}
Differentiating \cref{x_int,y_int} with respect to time and inserting \cref{andot,bndot} gives
\begin{align}
\label{xcdot}
\dot{\xc} &= \frac{\pi}{\Area} \int_{r_0}^{1/2} r \hat{u}_0(r,t) b_1(r,t) \, dr
- \frac{\pi}{\Area}  \left(r^2 \pd{a_1}{r} - r a_1 \right) \Big|_{r_0}^{r_1}, \\
\label{ycdot}
\dot{\yc} &= - \frac{\pi}{\Area}  \int_{r_0}^{1/2} r \hat{u}_0(r,t) a_1(r,t) \, dr
- \frac{\pi}{\Area}  \left(r^2 \pd{b_1}{r} - r b_1 \right) \Big|_{r_0}^{r_1}.
\end{align}

Given that the conductive-state solution \cref{conductive} takes the form of a Laurent polynomial, we consider a Laurent expansion of the variables $\hat{u}_0(r,t), \, a_1(r,t),\, b_1(r,t)$. Truncating each series to the lowest order capable of satisfying BCs \eqref{abinner}--\eqref{uvh_bc} gives
\begin{align}
\label{uu}
\hat{u}_0(r,t) &= C(t) (r-r_0) \left(1-2r \right) r^{-1},\\
\label{a1}
a_1(r,t) &=  \frac{1}{2} A(t) (2r-1) \left(1 - 2 r_0^2 r^{-1} \right), \\
\label{b1}
b_1(r,t) &= -\frac{1}{2} + \frac{1}{2} B(t) (2r-1) \left(1 - 2 r_0^2 r^{-1} \right).
\end{align}
where $(A,B,C)$ are time-dependent coefficients. We note that setting $A(t) = C(t) = 0$, $B(t) = -(4r_0^2+1)^{-1}$ recovers the conductive-state solution, \cref{conductive}, exactly.

Inserting \cref{uu,a1,b1} into \cref{L_int,x_int,y_int} gives linear relationships between $(L,\, \xc,\, \yc)$ and $(A,\, B,\, C)$:
\begin{align}
\label{LC}
L(t) &= \frac{(1-2 r_0)^2}{12} \, C(t) .\\
\label{COMAB1}
 \xc(t) &= \frac{(1-2r_0)^2 (1 + 6r_0 + 16r_0^2)}{48(1+2 r_0)}  \, A(t), \\
\label{COMAB2}
\yc(t) &=  \frac{1 + 2r_0 + 4 r_0^2}{12(1+2r_0)}
+ \frac{(1-2r_0)^2 (1 + 6r_0 + 16r_0^2)}{48(1+2 r_0)}  \, B(t) .
\end{align}

Evaluating the right-hand-sides of \cref{Ldot1,xcdot,ycdot} using \cref{uu,a1,b1,LC,COMAB1,COMAB2} gives the dynamical system \cref{Ldot,xdot,ydot} that is the main focus of this article.
The coefficients $\alpha,\, \beta,\, y_0,\, y_1$, and $k$ can be found by analytical integration (accelerated by symbolic programming). Each of these coefficients depends on $r_0$ only as given by 
\begin{align}
\label{alpha}
&\alpha = \frac{48}{(1-2r_0)^2}, \quad
\beta = \frac{48 (1+4r_0^2)}{(1-2r_0)^2 (1+6r_0+16r_0^2)}, \quad y_0 =  \frac{1+12r_0^2}{16 (1+4r_0^2)}, \\[5pt]
\label{keq}
&k = 24 \frac{(1-2r_0) 
(1 -6r_0 -4r_0^2 -88r_0^3 + 32r_0^4) 
-96 r_0^3 \ln{(2r_0)}} {(1-2r_0)^5 (1+6r_0+16r_0^2)}, \\[5pt] \label{y1}
&y_1 = \frac{(1-4r_0^2)
(1 -8r_0 -224r_0^3 -80r_0^4) - 192r_0^3(1+2r_0+4r_0^2)\ln{(2r_0)}}{24 (1-4r_0^2) 
(1 -6r_0 -4r_0^2 -88r_0^3 + 32r_0^4) 
-2304 (1+2r_0)r_0^3 \ln{(2r_0)}}.
\end{align}

\section{Stability analysis of the ODE system}\label{appD}

The fixed points of \cref{Ldot,xdot,ydot} are obtained by setting the right-hand-sides equal to zero. There can be up to three fixed points, given by:
\begin{enumerate}
    \item The conductive state:
    \begin{equation}
        \label{CondState}
        L =0, \quad \xc = 0, \quad \yc = y_0.
    \end{equation}
    \item  The circulating states:
    \begin{equation}
        \label{CircState}
        L = \pm L_1, \quad \xc = \mp\frac{\alpha}{\Ra} L_1, \quad  \yc = y_1 + \frac{\alpha\beta}{k \Ra},
    \end{equation}
    where 
    \begin{equation}
        L_1 = \pm \frac{\beta}{k} \sqrt{ \frac{k \Ra}{\alpha \beta} \dy -1 }.
    \end{equation}
\end{enumerate}
We note that the circulating states only exist if $\Ra \geq \alpha \beta / (k \dy) = \Ra^*_1$.

The Jacobian of \cref{Ldot,xdot,ydot} is
\begin{equation}
\label{Jacobian}
J(L,\xc,\yc) = \left[ {\begin{array}{ccc}
   -\alpha\,\Pra& -\Ra\,\Pra & 0 \\
   -k(\yc-y_1)  & -\beta    & -kL \\
   k\xc         & kL    & -\beta
  \end{array} } \right].
\end{equation}

Evaluating the Jacobian at fixed point \cref{CondState}, one can show that all eigenvalues are negative if $\Ra < \Ra^*_1$, thereby confirming the conductive state is stable when $\Ra < \Ra^*_1$. Above $\Ra^*_1$, \cref{CondState} becomes unstable and the two circulating states given by \cref{CircState} appear, indicating a pitchfork bifurcation. The circulating states are stable provided that $\Ra^*_1 < \Ra < \Ra^*_2$, with $\Ra^*_2$ defined in \cref{ra12}. Above $\Ra^*_2$, the real part of the complex eigenvalues become positive, rendering the circulating states unstable.

\section{High-$\Ra$ LSC reversal frequency}\label{appE}

At very high $\Ra$, order reemerges in the system and the large-scale dynamics become nearly periodic. The dominant frequency $f^*$ of the LSC reversals can be obtained by the pendulum equivalence of \cref{Ldot,xdot,ydot}. In this correspondence, the gravitational constant is $\geff = k l^2 \, \Ra \, \Pra$, which gives a small-amplitude frequency of $\sqrt{k l \, \Ra \, \Pra}/(2\pi)$. The oscillation amplitude, however, is not small, which implies that $f^*$ depends on both the pendulum length $l$ and the maximum swing angle $\phim$, both of which can be estimated through an energy balance.

Multiplying \cref{xdot} by $\xc$, \cref{ydot} by $\yc$, and adding gives the exact relation
\begin{equation}
\td{}{t}{l^2} = -2 \beta l^2 + 2\beta \dy (Y-y_1).
\end{equation}
Assuming periodicity implies that the time average of $dl^2/dt$ vanishes, giving
\begin{align}
\label{meanell2}
\mean{l^2} = \dy \mean{Y - y_1} ,
\end{align}
where $\mean{\cdot}$ indicates a time average.

Consider the effective energy of the pendulum system
\begin{equation}
\label{energy}
\Eeff = \frac{1}{2} k L^2 + \Ra \Pra (\yc-y_1) .
\end{equation}
Here, the terms on the right-hand-side  represent the kinetic and potential energy of the mechanical pendulum respectively.
Taking a time derivative and using \cref{Ldot,ydot}, gives the energy law
\begin{equation}
\label{Edot}
\dot{E}_{\text{eff}} = -\alpha \Pra \, k L^2 + \beta \Ra \Pra (y_0-\yc) .
\end{equation}
The first term above represents energy dissipation due to damping while the second term represents positive energy injected into the system from the external driving. The assumption of periodicity implies $\mean{\dot{E}_{\text{eff}}} = 0$, which gives
\begin{equation}
\label{meanL}
k \alpha \mean{L^2} = \Ra\, \beta \mean{y_0 - \yc} .
\end{equation}
Meanwhile, directly averaging \cref{energy} gives
\begin{equation}
\label{Eavg}
\mean{\Eeff} = \frac{1}{2} k \mean{L^2} + \Ra \Pra \mean{\yc-y_1} .
\end{equation}

At the bottom of the swing, $Y = y_1 - l$, $L = L_{\max{}}$, the energy is
\begin{equation}
\label{Ebot}
\Ebot = \frac{1}{2} k \Lmax^2 - \Ra \Pra \, l .
\end{equation}
Due to periodicity, $\mean{L^2} = \Lmax^2/m$, where $m$ is a constant.
Although energy is not strictly conserved, it is conserved on average for periodic dynamics. We therefore make the assumption of nearly constant energy, $\Ebot = \mean{E}$ in order to solve for $l$ and $\phim$. Setting \cref{Eavg} equal to \cref{Ebot} and using \cref{meanell2,meanL} gives
\begin{equation}
\label{leq}
l = \dy \left( \frac{(m-1) \beta}
{(m-1)\beta + 2 \alpha \Pra} \right) .
\end{equation}
At the apex, $\phi = \phim$ and $\dot{\xc}=\dot{\yc}=0$. \Cref{xdot,ydot} then simplify to $ \xc^2 = (y_0-\yc)(\yc-y_1)$. As $\xc = l \sin\phi$ and $\yc = y_1 - l \cos \phi$, we have  $l = -\dy \cos \phim$, thus providing the value of $\phim$ once $l$ is given by \cref{leq}. Based on \cref{fig1}(e), we choose the value $m=2.5$ as the midpoint of a sinusoidal ($m=2$) and a triangular ($m=3$) waveform. 

With the values of $l$ and $\phim$ determined, the frequency of (large-amplitude) pendulum oscillation, and therefore the frequency of regular LSC reversals, is given by \cref{f*} in the text. We note that the parameters of \cref{fig4}, $\Pr = 4$ and $r_0 = 0.4$, give values $l = 0.005$ and $\phim = 1.62$.

Converting \cref{f*} to a dimensional frequency gives 
\begin{equation*}
    F^* = \frac{\kappa}{h^2}f^* = c N^*,
\end{equation*}
where $N^* = \sqrt{\beta_T \Delta Tg/h}$ is the Brunt–V\"ais\"al\"a frequency and
\begin{align*}
    c =  \frac{\sqrt{kl}}{ 4 \, K(\sin^2 ({\phim}/{2})) }
\end{align*}
is a constant that depends only on $r_0$ and $\Pra$. In particular, $c$ is independent of the Rayleigh number.

\bibliographystyle{jfm}
\bibliography{manuscript}

\end{document}